# Properties of Gamow-Teller and charge-exchange giant spin-monopole resonances in medium-heavy closed-shell parent nuclei: a semi-microscopic description.

V.I. Bondarenko[1], M.H. Urin[2]

[1] *Shubnikov Institute of Crystallography, Federal Research Center "Crystallography and Photonics," Russian Academy of Sciences, Moscow, 119333, Russia;*

[2] *National Research Nuclear University "MEPhI", Moscow,115409 , Russia*

The basic version of the semi-microscopic particle-hole dispersive optical model is adopted and implemented to describe main properties of the Gamow-Teller and charge-exchange giant spin-monopole resonances in medium-heavy closed-shell parent nuclei. Calculation results obtained for $^{48}$Ca, $^{90}$Zr, $^{132}$Sn, and $^{208}$Pb are compared with available experimental data.

## I. Introduction

Main properties of giant resonances (GRs) related to high-energy excitations of the particle-hole type in medium-heavy nuclei are described by a number of characteristics. These are: the energy-averaged strength function and projected (one-body) transition density both related to an appropriate external field (one-body probing operator), and probabilities of direct one-nucleon decay.

The mentioned characteristics can be evaluated within a model, in which the main relaxation modes of particle-hole (p-h) states associated with GRs are together taken into account. The relaxation modes include Landau damping, coupling the p-h states to the single-particle (s-p) continuum and to many-quasiparticle configurations (the spreading effect). The recently developed particle-hole dispersive optical model (PHDOM) [1] can be considered as such model. Being formulated in a rather general form for medium-heavy closed-shell nuclei, the PHDOM is an extension of the continuum-random-phase-approximation (cRPA) standard [2] and nonstandard [3] versions to taking into account the spreading effect. Within the PHDOM, Landau damping and coupling to s-p continuum are described microscopically in terms of a mean field and p-h interaction, while the spreading effect is treated phenomenologically in terms of the respective energy-averaged p-h self-energy term. The imaginary part of this term determines the real part via a microscopically-based dispersive relationship. For the above reasons, the PHDOM is related to semi-microscopic models.

In recent years, the basic PHDOM version has been adopted and then implemented for describing properties of some isoscalar and isovector GRs in medium-heavy closed-shell nuclei (see, e.g., Ref. [4] and references therein). In these implementations, a realistic phenomenological (of the Woods-Saxon type) partially self-consistent mean field (with parameters taken from independent data), and Landau-Migdal p-h interaction have been exploited. Two self-consistency conditions used are related to isospin-symmetry and translation invariance of the model Hamiltonian. Among references given in Ref. [4], we point out Ref. [5], where the PHDOM-based description of Isobaric Analog Resonance and its overtone (i.e., Isovector Giant Monopole Resonance in the $\beta^{(-)}$-channel) has been proposed. In the following, we use a similarity in describing characteristics of the mentioned $0^+$-GRs and Gamow-Teller Resonance (GTR), its overtone, (i.e., Isovector Giant Spin-Monopole Resonance in the $\beta^{(-)}$-channel (IVGSMR$^{(-)}$)). In the present work, the main characteristics of these spin-flip $1^+$-GRs (together with the isobaric partner of IVGSMR$^{(-)}$ – IVGSMR$^{(+)}$) in the $^{48}$Ca, $^{90}$Zr, $^{132}$Sn, and $^{208}$Pb parent nuclei are evaluated within the properly adopted PHDOM basic version. In applying to $^{208}$Pb, preliminary results of this study are given in Ref. [6]. We note also studies of Refs. [7, 8], where, in fact, "pole" versions of PHDOM have been exploited in describing the mentioned $1^+$-GRs. Concluding this Introduction, we note possibilities to extend the basic PHDOM version. One of possibilities consists in taking into account tensor correlations in GR formation. In applying to the charge-exchange $1^+$-GRs in $^{208}$Pb parent nucleus, this problem has been recently studied within the properly extended PHDOM version in Ref. [9].

In Sect. II, the main model relations are presented. Input quantities and choice of model parameters are described in Sect. III. Sect. IV contains calculation results and discussions. Summary and conclusive remarks are given in Sect. V.

## II. Main model relations

The PHDOM basic version has been originally formulated in terms of the energy-averaged p-h Green function (p-h effective propagator). However, for evaluating the GR characteristics listed in Introduction, it is simpler to use the effective-field method introduced in nuclear physics by Migdal [10]. Within the model, the effective field is defined via a convolution of the p-h Green function with the related external field (the s-p probing operator), leading to excitation of a given GR. As a result, the integral equation for the effective field follows from the Bethe-Goldstone-type equation for the p-h Green function.

We start presenting the main model relations from the isovector component of Landau-Migdal forces responsible for long-range correlations and leading to formation of isovector GRs :

$$F_{IV}(x_1, x_2) = (F' + G'\vec{\sigma}_1\vec{\sigma}_2)\vec{\tau}_1\vec{\tau}_2\delta(\vec{r}_1 - \vec{r}_2). \qquad (1)$$

Here, $(F', G') = (f', g') * 300$ MeV Fm$^3$ are the well-known Landau-Migdal parameters. In particular, the (dimensionless) parameter $f'$, determining the symmetry potential via the isospin self-consistency condition, might be related to mean-field parameters (see, e.g., Ref. [4]). Searching the parameter $g'$ from the experimental GTR energy becomes a certain trend (see, e.g., Ref. [11]).

Let $V_M^{(-)}(x) = \tau^{(-)}\sigma_M V(r)$ be the s-p external fields, leading to excitation of the monopole spin-flip GRs in the $\beta^{(-)}$-channel ($\tau^{(-)}$ and $\sigma_M$ are the respective

Pauli matrixes). For describing GTR and IVGSMR$^{(-)}$, the radial part of the external fields might be taken as $V_{\text{GT}}(r) = 1$ and $V_{\text{SM}}(r) = r^2 - \eta$, respectively. The parameter $\eta$ is determined by the condition of minimal GTR excitation by probing operator (Sect. III). In neglecting tensor correlations, the effective fields $\tilde{V}_M^{(-)}(x, \omega)$, which are different from related external fields due to the p-h interaction of Eq. (1), have the same spin-angular dependence [9]. Hereafter, $\omega = E_x + Q^{(-)}$, where $E_x$ is the excitation energy of the isobaric nucleus and $Q^{(-)}$ is the difference of ground-state energies of the isobaric and parent nuclei, is the isobaric-nucleus (*Z+1, N-1*) excitation energy counted off from the parent-nucleus (*Z, N*) ground-state energy. The effective-field radial part obeys the equation obtained after separation of spin-angular and isospin variables in the above-mentioned integral equation [6, 9]:

$$\tilde{V}^{(-)}(r, \omega) = V(r) + \frac{G'}{2\pi r^2} \int A_\sigma^{(-)}(r, r', \omega)\, \tilde{V}^{(-)}(r', \omega) dr'. \qquad (2)$$

Here, $(4\pi rr')^{-2} A_\sigma^{(-)}(r, r', \omega)$ is the spin-monopole radial part of the "free" p-h propagator in the $\beta^{(-)}$–channel, $A^{(-)}(x, x', \omega)$. Being the PHDOM key quantity, the "free" p-h propagator is related to the model of non-interacting and independently damping p-h excitations. A rather cumbersome expression (multiplied by $4\pi$) for the monopole radial part of $A^{(-)}(x, x', \omega)$ is given in detail in Ref. [5]. The following substitution of squared kinematic factors in this expression,

$$t^2_{(\pi)(\nu)} = (2j_\pi + 1)\delta_{(\pi)(\nu)} \rightarrow \left(t^\sigma_{(\pi)(\nu)}\right)^2 = \frac{1}{3}\langle(\pi)\|\sigma\|(\nu)\rangle^2, \qquad (3)$$

allows one to get the explicit expression for the quantity $A_\sigma^{(-)}(r, r', \omega)$ in Eq. (2). Both expressions contain: the occupation numbers $n_\mu$ for proton ($\mu = \pi$) and neutron ($\mu = \nu$) levels with $\mu$ being the set of single-particle quantum numbers $n_{r,\mu}$, $j_\mu$, $l_\mu$ (($\mu$) = $j_\mu$, $l_\mu$); the bound-state energies $\varepsilon_\mu$ and radial wave functions $r^{-1}\chi_\mu(r)$; and proton and neutron optical-model-like Green functions of the radial Schrodinger equations, in which the mean field has an additional term proportional to the strength of the p-h self-energy term responsible for the spreading effect, ($-iW(E_x) + P(E_x)$). The proton radial Schrodinger equation determines the optical-model-like proton radial continuum-state wave functions, $r^{-1}\chi_{\varepsilon>0,(\pi)}(r)$, used below, having the standing-wave asymptotic behavior and obeying the δ-function energy normalization in the limit $W = P = 0$.

The effective-field radial part of Eq. (2) determines the following energy-averaged GTR and IVGSMR$^{(-)}$ characteristics considered for a wide excitation energy interval:

(i) the strength function, $S_V^{(-)}(\omega)$,

$$S_V^{(-)}(\omega) = -\frac{1}{\pi} \mathrm{Im} \int V(r) A_\sigma^{(-)}(r, r', \omega) \tilde{V}^{(-)}(r', \omega) dr dr'; \qquad (4)$$

(ii) the radial part of the projected transition density, $r^{-2}\rho_V^{(-)}(r, \omega)$,

$$r^{-2}\rho_V^{(-)}(r,\omega) = -\frac{2 Im\,\tilde{V}^{(-)}(r,\omega)}{G'\sqrt{S_V^{(-)}(\omega)}}; \qquad (5)$$

(iii) the partial strength function and related branching ratio for direct one-proton decay accompanied by population of the product-nucleus neutron-hole state $\nu^{-1}$, $S_{V,\nu}^{(-),\uparrow}$ and $b_{V,\nu}^{(-),\uparrow}$, respectively,

$$S_{V,\nu}^{(-),\uparrow}(\omega) = \sum_{(\pi)} n_\nu \left(t_{(\pi)(\nu)}^\sigma\right)^2 \left|\int \chi_{\varepsilon=\varepsilon_\pi+\omega,(\pi)}^*(r)\tilde{V}^{(-)}(r,\omega)\chi_\nu(r)dr\right|^2 \qquad (6)$$

and

$$b_{V,\nu}^{(-),\uparrow}(\delta^{(-)}) = \int_{\delta^{(-)}} S_{V,\nu}^{(-),\uparrow}(\omega)d\omega \Big/ \int_{\delta^{(-)}} S_V^{(-)}(\omega)d\omega. \qquad (7)$$

In Eq. (7), $\delta^{(-)}$ is the excitation-energy interval that includes the considered GR.

The main characteristics of IVGSMR$^{(+)}$ (spin-monopole $1^+$-GR in the $\beta^{(+)}$-channel), namely, the strength function $S_V^{(+)}(\omega)$, the radial part of the projected transition density $r^{-2}\rho_V^{(+)}(r,\omega)$, the partial strength function and related branching ratio for direct one-neutron decay accompanied by population of the product-nucleus proton-hole state $\pi^{-1}$, $S_{V,\pi}^{(+),\uparrow}$ and $b_{V,\pi}^{(+),\uparrow}$, respectively, are determined by the radial part of the effective field, $\tilde{V}^{(+)}(r,\omega)$, corresponding to the external field $V_M^{(+)}(x) = \tau^{(+)}\sigma_M V(r)$. The equation for the mentioned effective field and the expressions for the above-listed characteristics of IVGSMR$^{(+)}$ can be obtained by the interchange (π) ↔ (ν) in, Eq. (2) and relations (4) – (7), respectively. In such a case, $\omega = E_x + Q^{(+)}$ is the isobaric-nucleus (Z-1, N+1) excitation energy counted off from the parent-nucleus (Z, N) ground-state energy ($E_x$ is the excitation energy of the isobaric nucleus, $Q^{(+)}$ is the difference of ground-state energies of the isobaric and parent nuclei). The statements given in this paragraph are formally related also to weak Gamov-Teller-type excitations in the $\beta^{(+)}$ – channel. These excitations are taken into consideration in evaluation of the respective sum rule (see below).

We now provide comments to the above-given expressions for main characteristics of the considered charge-exchange $1^+$-GRs:

1) The strength functions $S_V^{(\mp)}(E_x)$ obey the non-energy-weighted sum rule, (NEWSR$_V$):

$$\text{NEWSR}_V = \int_0^\infty S_V^{(-)}(E_x)dE_x - \int_0^\infty S_V^{(+)}(E_x)dE_x = 4\pi \int_0^\infty V^2(r)n^{(-)}(r)r^2 dr, \quad (8)$$

where $n^{(-)}(r)$ is the neutron-excess density in the parent nucleus. In particular, the well-known Ikeda sum rule for GT excitation (NEWSR$_{GT} = N - Z$) follows from this relation, which is also valid within cRPA ($W = P = 0$). To verify the strength function calculations, we compare with unity the value of the fraction parameter $x_V^*$ defined for a large cut-off excitation energy $E_x^*$:

$$x_V^* = \left(\int_0^{E_x^*} S_V^{(-)}(E_x)dE_x - \int_0^{E_x^*} S_V^{(+)}(E_x)dE_x\right)/\text{NEWSR}_V. \quad (9)$$

This relation makes it reasonable to consider the reduced strength functions

$$y_V^{(\mp)}(E_x) = S_V^{(\mp)}(E_x)/\text{NEWSR}_V, \tag{10}$$

determining the fraction parameters $x_V^{(\mp)}(\delta) = \int_{E_{x_2}}^{E_{x_1}} y_V^{(\mp)}(E_x)\, dE_x$ for a given excitation-energy interval $\delta = E_{x_1} \div E_{x_2}$.

2) Calculations of the (one-dimensional) radial part of the projected one-body transition densities $\rho_V^{(\mp)}(r,\omega)$ performed in accordance with Eq. (5) might be verified by the relations

$$S_V^{(\mp)}(E_x) = \left(\int_0^\infty V(r)\rho_V^{(\mp)}(r,E_x)\, dr\right)^2. \tag{11}$$

These relations follow from the definition of the projected transition densities, which are determined by a convolution of the energy-averaged double transition densities with the respective probing operator (see, e.g., Ref. [4]). The one-body transition densities considered within cRPA are independent of $V$. The relation (11) allows one to expect that transition densities $\rho_V^{(\mp)}(r,E_x)$ are mainly proportional to $\left[S_V^{(\mp)}(E_x)\right]^{1/2}$. In other words, the reduced transition densities $\bar{\rho}_V^{(\mp)}(r,E_x) = \rho_V^{(\mp)}(r,E_x)/\left[S_V^{(\mp)}(E_x)\right]^{1/2}$ are expected to be slightly dependent of $E_x$.

3) The total branching ratios for direct one-nucleon decay

$$b_{V,tot}^{(-),\uparrow} = \sum_\nu b_{V,\nu}^{(-),\uparrow}, \quad b_{V,tot}^{(+),\uparrow} = \sum_\pi b_{V,\pi}^{(+),\uparrow} \tag{12}$$

determine the branching ratios for statistical (mainly neutron) decay of GTR and IVGSMR$^{(\mp)}$, $b_V^{(\mp),\downarrow} = 1 - b_{V,tot}^{(\mp),\uparrow}$. In the absence of the spreading effect (i.e., within cRPA), when $\sum_\nu S_{V,\nu}^{(-),\uparrow}(\omega) = S_V^{(-)}(\omega)$ and $\sum_\pi S_{V,\pi}^{(+),\uparrow}(\omega) = S_V^{(+)}(\omega)$ (the unitary condition), $b_{V,tot}^{(\mp),\uparrow} = 1$ and, naturally, $b_V^{(\mp),\downarrow} = 0$.

### III. Input quantities. Model parameters

For describing within PHDOM main characteristics of GTR and IVGSMR$^{(\mp)}$ in medium-heavy doubly-closed-shell parent nuclei, the following input quantities and model parameters are used:

1) The isovector part of the Landau-Migdal p-h interaction of Eq. (1). For the parent nuclei under consideration, the parameter $g'$ of this interaction is adjusted to describe, in model calculations of the GT strength function, the observable GTR peak energy $E_{x,peak}^{exp}$. Apart from possibilities to describe characteristics of IVGSMR$^{(\mp)}$ in these nuclei, such a procedure allows us to see how the $g'$ value is changed with $A$. The choice of the $f'$ value is explained in the next point. The deduced values of the Landau-Migdal parameters are given in Table I.

2) The realistic (Woods-Saxon type) phenomenological partially self-consistent mean field described in details in Refs. [12, 4]. The mean field contains: phenomenological isoscalar central and spin-orbit parts, $U_0(r)$ and $U_{ls}(r)$, with intensities $U_0$ and $U_{ls}$, respectively; the symmetry potential $U_{sym} = \frac{1}{2}v(r)\tau^{(3)}$ and the mean Coulomb field $U_C(r)$, both calculated self-consistently via the neutron-excess and proton densities ($n^{(-)} = n^n - n^p$ and $n^p$), respectively. Due to the isospin self-consistency condition, $v(r) = 2F'n^{(-)}(r)$, the Landau-Migdal parameter $f'$ might be related to mean-field parameters. The set of mean-field parameters contains the intensities $U_0$ and $U_{ls}$, Woods-Saxon size and diffuseness parameters ($r_0$ and $a$, respectively), and Landau-Migdal parameter $f'$. Employing the method used in Refs. [12, 4], we found the above-listed parameters for doubly-closed-shell nuclei $^{48}$Ca, $^{132}$Sn, $^{208}$Pb by reproducing the observable single-quasiparticle spectra in the respective even-odd and odd-even nuclei. The obtained values[*] are used as a base to get the mean-field parameters for $^{90}$Zr by means of the interpolation procedure described in Ref. [13]. The mean-field parameters used in calculations of characteristics of GTR and IVGSMR$^{(\mp)}$ in parent nuclei under consideration are given in Table I.

3) The imaginary part of the strength of the energy-averaged p-h self-energy term responsible for the spreading effect, $W(E_x)$ (Sect. II). Following previous PHDOM implementations (Ref. [4] and references therein), we take this phenomenological quantity as the universal three-parametric function of excitation energy (counted off the compound-nucleus ground-state energy):

$$W(E_x) = \begin{cases} 0, & E_x < \Delta \\ \alpha(E_x - \Delta)^2 / [1 + (E_x - \Delta)^2/B^2], & E_x \geq \Delta. \end{cases} \quad (13)$$

---

[*] These parameters are found in collaboration with M.L. Gorelik and B.A.Tulupov

Here the "spreading" parameters $\alpha$, $\Delta$ and $B$ can be called as the strength, gap and saturation parameters, respectively. As in previous PHDOM implementations, we take for nuclei under consideration the gap parameter as universal quantity. The parameters $\alpha$ and $B$ are adjusted to describe within the model the observable total width of GTR, $\Gamma_{exp}$, in the parent nuclei under consideration. Obtained values of the adjusted "spreading" parameters are also given in Table I. The real part of the strength of the energy-averaged p-h self-energy term, $P(E_x)$, is determined by the imaginary part, $W(E_x)$, via the microscopically-based dispersive relation [1]. For $W(E_x)$ of Eq. (13), the expression for $P(E_x)$ is rather cumbersome and can be found in Ref. [14].

4) The parameter $\eta$ in the expression for the SM probing-operator radial part, $V_{SM}(r)$ (see Sect. II), is found from the condition of minimal GTR excitation by the respective external field. This condition might be taken in the form: $\min \int S_{SM}^{(-)}(E_x) dE_x$, where integration is performed over the GTR region. The values of parameter $\eta$ obtained for nuclei under consideration are given in Table I.

TABLE I. The model parameters used in evaluation of GR characteristics for nuclei under consideration. (Notations are given in the text). The parameters $r_0 = 1.21$ fm and $\Delta = 3$ MeV are taken as universal quantities.

| Nucleus | $U_0$, (MeV) | $U_{ls}$, (MeV fm$^2$) | $a$, (fm) | $f'$ | $g'$ | $\eta$, (fm$^2$) | $\alpha$, (MeV$^{-1}$) | $B$, (MeV) |
|---|---|---|---|---|---|---|---|---|
| $^{48}$Ca | 54,34 | 32,09 | 0,58 | 1,13 | 0,85 | 17,74 | 0,26 | 5,27 |
| $^{90}$Zr | 55,06 | 34,93 | 0,61 | 1,05 | 0,69 | 23,67 | 0,52 | 5,27 |
| $^{132}$Sn | 55,53 | 35,98 | 0,63 | 1,00 | 0,69 | 30,85 | 0,33 | 5,27 |
| $^{208}$Pb | 55,74 | 33,35 | 0,63 | 0,98 | 0,71 | 40,16 | 0,27 | 4,88 |

**IV. Characteristics and parameters of GTR, IVGSMR$^{(\mp)}$**

In this section we present the main characteristics and parameters of GTR, IVGSMR$^{(\mp)}$ calculated within PHDOM for the $^{48}$Ca, $^{90}$Zr, $^{132}$Sn and $^{208}$Pb parent nuclei, using the model parameters listed in Table I. The GR strength functions $S_V^{(\mp)}(E_x)$ (a) and $y_V^{(\mp)}(E_x)$ (b) calculated by employing Eqs. (4) and (10), respectively, in a large excitation-energy interval $\delta^* = 0 \div E_x^*$ are shown in Figs 1-3 ($E_x^* + Q = 80$ MeV).

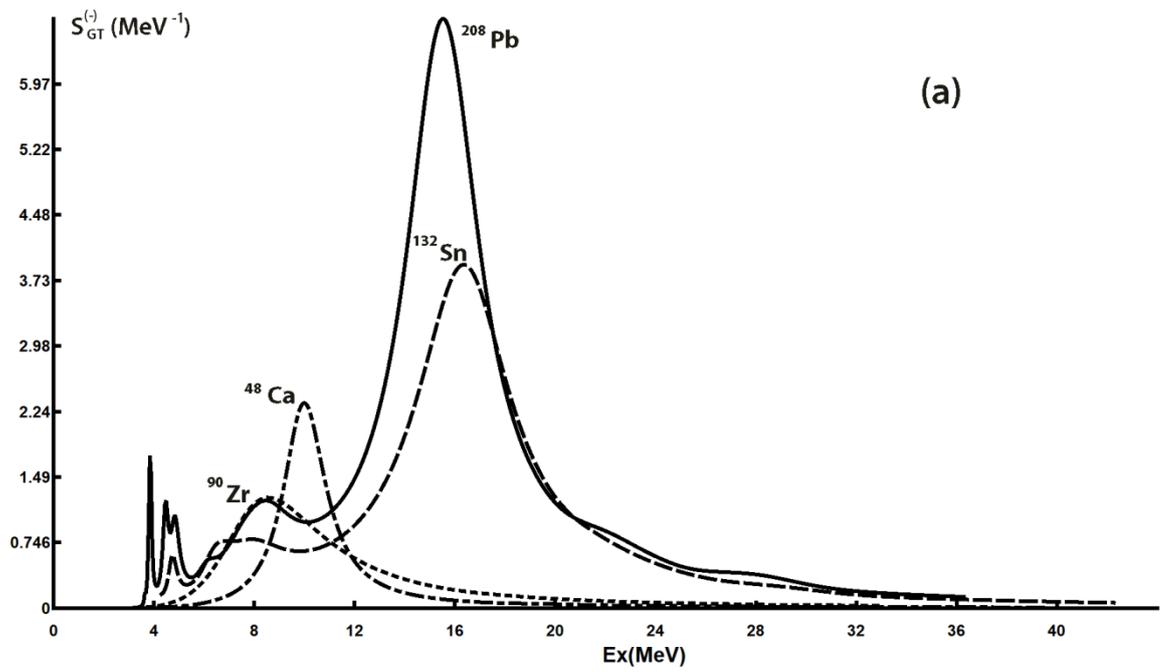

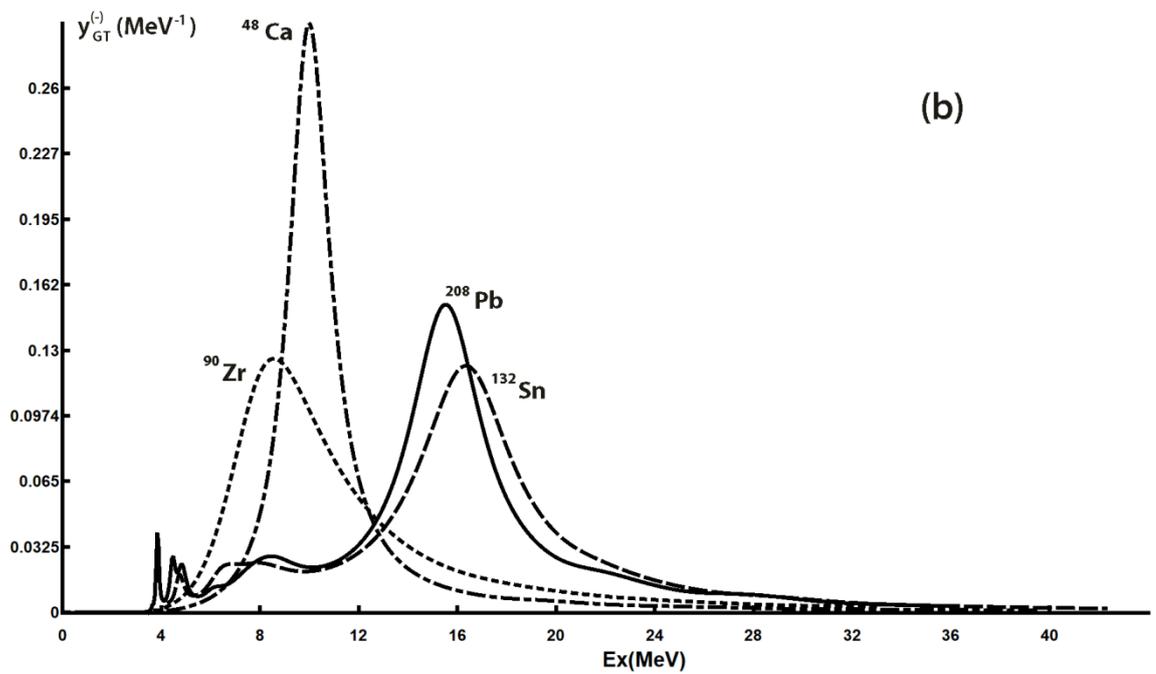

FIG.1 The GTR strength functions calculated within PHDOM for parent nuclei under consideration.

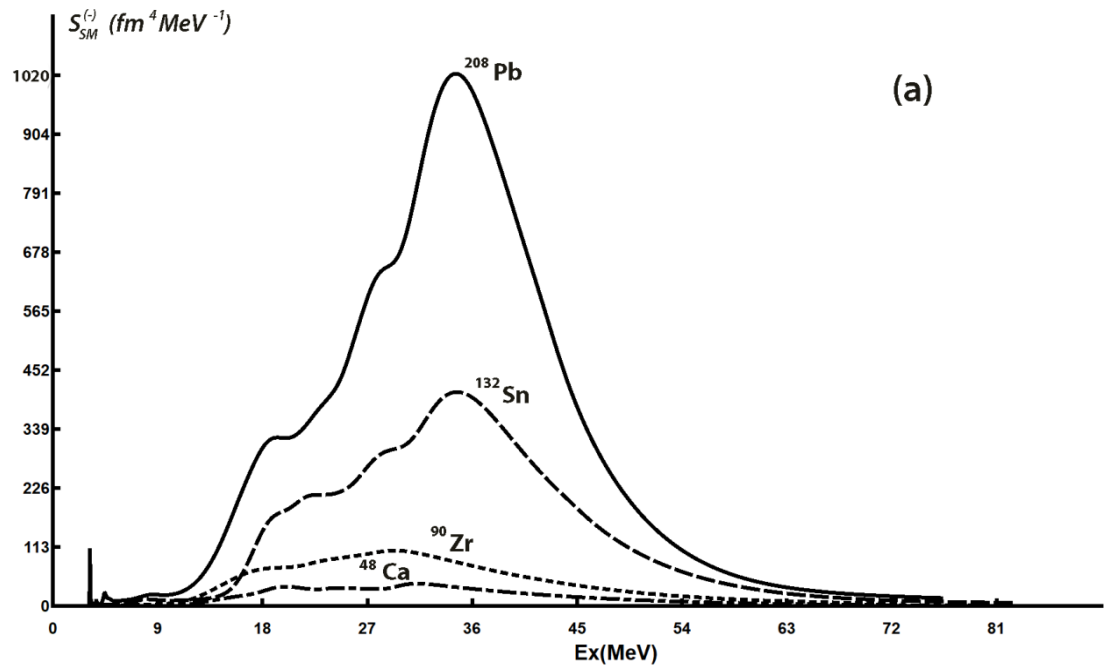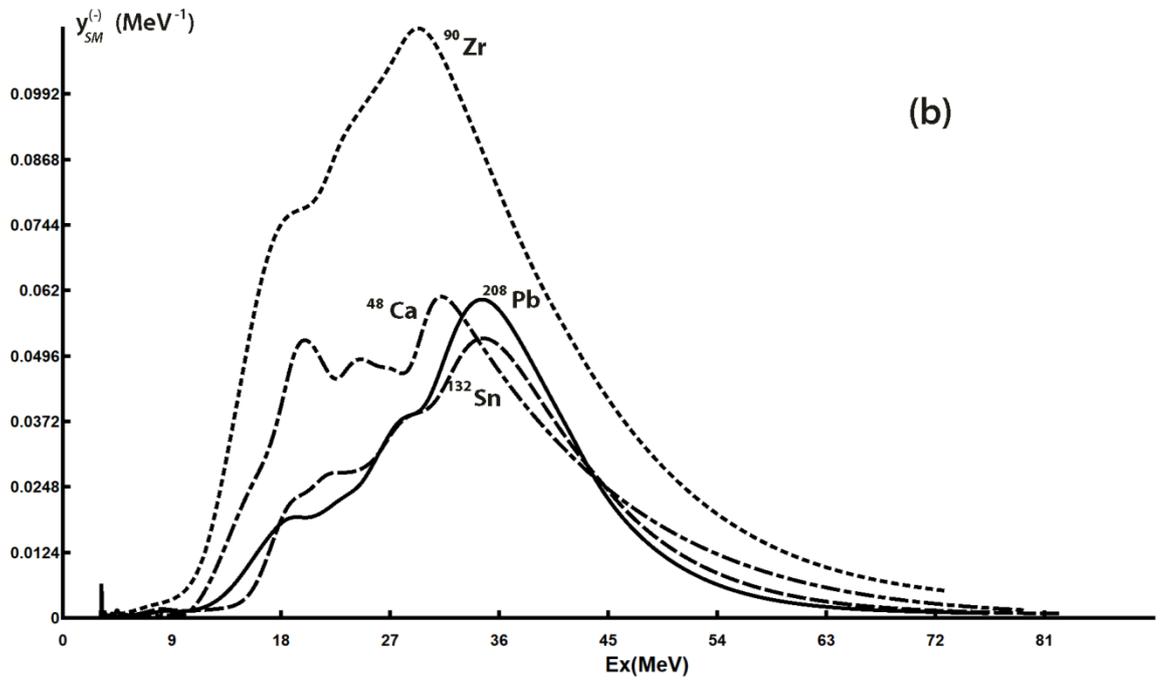

FIG. 2 The same as in Fig. 1, but for IVGSMR$^{(-)}$.

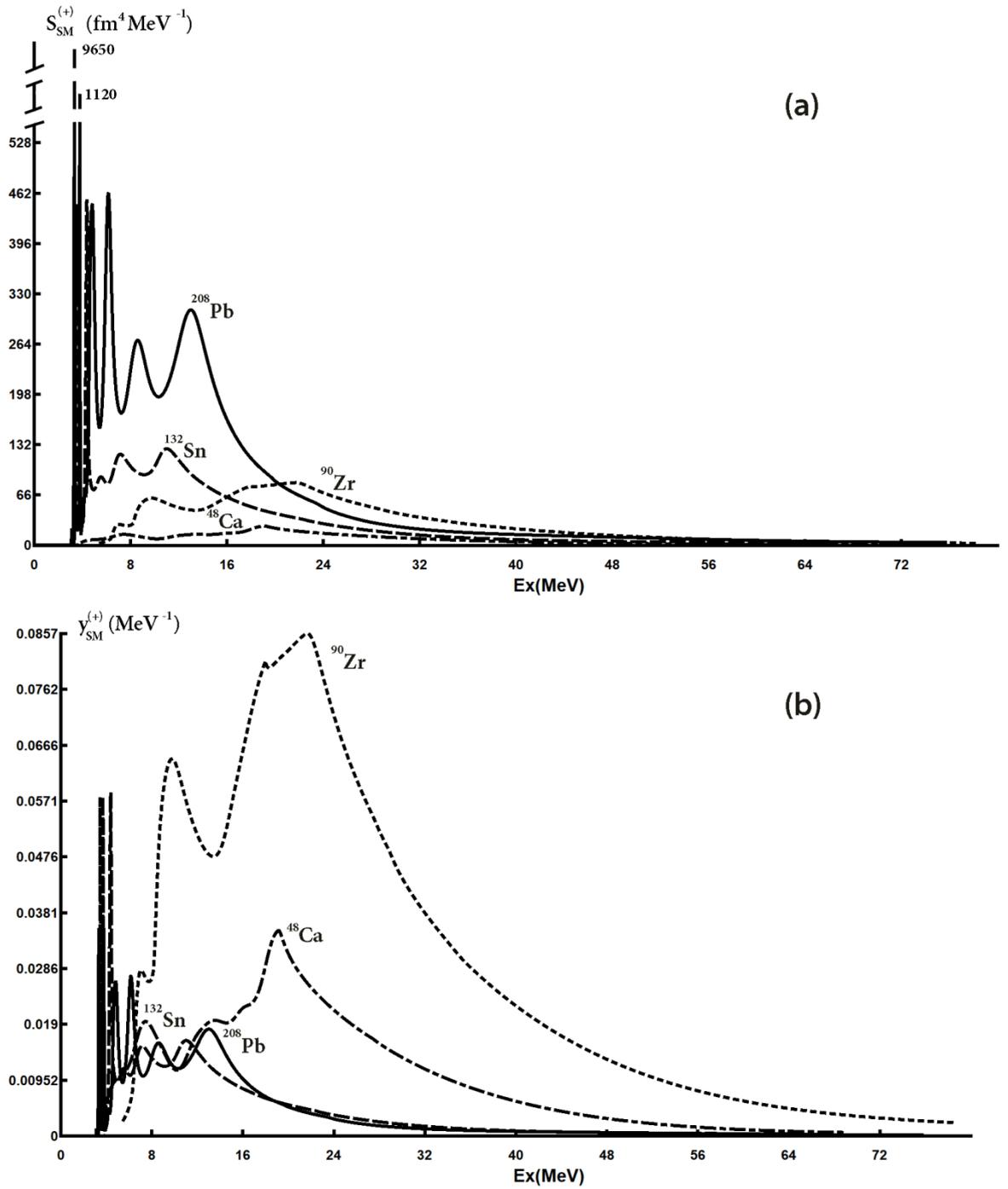

FIG. 3 The same as in Fig. 1, but for IVGSMR$^{(+)}$.

The calculated strength functions allow us to evaluate the following parameters of considered GRs: the peak energy and total width, $E_{x,peak}$ and $\Gamma$, respectively, (both found by means of the Lorentz-type parameterization of the strength function $S_V(E_x)$ near its main maximum); the fraction parameters $x_{max}^{(\mp)}$ defined for the excitation-energy interval $\delta_{max}^{(\mp)}$ in the vicinity of the strength-function main maximum, $x^{(\mp),*}$ defined for the excitation-energy intervals considered in Figs 1-3, and $x^* = x^{(-),*} - x^{(+),*}$. The above-listed parameters are given in Tables II – IV.

TABLE II. The GTR parameters evaluated within PHDOM for parent nuclei under consideration. The reference values $E_{x,peak}^{exp}$ and $\Gamma_{exp}$ are taken from Refs. [11, 13-17].

| Nucleus | $E_{x,peak}$, (MeV) | $\Gamma$, (MeV) | $\delta_{max}^{(-)}$, (MeV) | $x_{max}^{(-)}$, (%) | $x^{(-),*}$, (%) | $x^*$, (%) | $b_{tot}^{(-),\uparrow}$, (%) |
|---|---|---|---|---|---|---|---|
| $^{48}$Ca | 9.98 | 2.0 | 4.50 ÷ 14.50 | 78 | 89 | 88 | 0.6 |
| $^{90}$Zr | 8.76 | 4.88 | 6.10 ÷ 13.10 | 59 | 88 | 83 | 2.5 |
| $^{132}$Sn | 16.32 | 4.7 | 10.31 ÷ 24.81 | 75.6 | 98 | 97 | 1.9 |
| $^{208}$Pb | 15.50 | 3.7 | 10.44 ÷ 20.34 | 69.2 | 101 | 99 | 4.3 |

TABLE III. The IVGSMR$^{(-)}$ parameters evaluated within PHDOM for parent nuclei under consideration.

| Nucleus | $E_{x,peak}$, (MeV) | $\Gamma$, (MeV) | $x^{(-),*}$, (%) | $x^*$, (%) | $\delta_{max}^{(-)}$, (MeV) | $b_{tot}^{(-),\uparrow}$, (%) |
|---|---|---|---|---|---|---|
| $^{48}$Ca | 31.58 | 18.96 | 165 | 99 | 24.50 ÷ 40.50 | 75 |
| $^{90}$Zr | 26.99 | 26.17 | 301 | 92 | 14.10 ÷ 43.10 | 54 |
| $^{132}$Sn | 33.83 | 19.42 | 124 | 99 | 23.31 ÷ 54.31 | 51 |
| $^{208}$Pb | 34.60 | 16.80 | 124 | 95 | 18.34 ÷ 46.34 | 51 |

TABLE IV. The same as in Table III, but for IVGSMR$^{(+)}$.

| Nucleus | $E_{x,peak}$, (MeV) | $\Gamma$, (MeV) | $x^{(+),*}$, (%) | $x^*$, (%) | $\delta_{max}^{(+)}$, (MeV) | $b_{tot}^{(+),\uparrow}$, (%) |
|---|---|---|---|---|---|---|
| $^{48}$Ca | 19.10 | 16.22 | 66 | 99 | 17.84 ÷ 29,84 | 70 |
| $^{90}$Zr | 20.14 | 17.40 | 208 | 92 | 14.51 ÷ 41.51 | 59 |
| $^{132}$Sn | 11.06 | 8.26 | 25 | 99 | 9.65 ÷ 15.65 | 57 |
| $^{208}$Pb | 12.95 | 6.19 | 29 | 95 | 9.78 ÷ 17.78 | 44 |

The next main GR characteristic, which might be calculated within PHDOM, is the one-body (projected) transition density $\rho_V^{(\mp)}(r, E_x)$ of Eqs. (5), (11). The transition densities calculated at the peak energy of GTR, IVGSMR$^{(\mp)}$ in parent nuclei under consideration are shown in Figs. 4(a) – 6(a). As expected for main-tone and overtone GRs, the transition density exhibits node-less (for GTR) and one-node (for IVGSMR$^{(\mp)}$) radial dependence. The transition-density energy dependence is also the subject of interest in view of possibility to use the transition density for describing nuclear reactions of GR excitation. In Figs. 4(b) – 6(b), we show the reduced transition densities $\bar\rho_V^{(\mp)}(r_{max}, E_x)$ (Sect. II) calculated at the maximum of the transition densities $\rho_V^{(\mp)}(r, E_{x,peak})$ presented in Figs 4(a)-6(a).

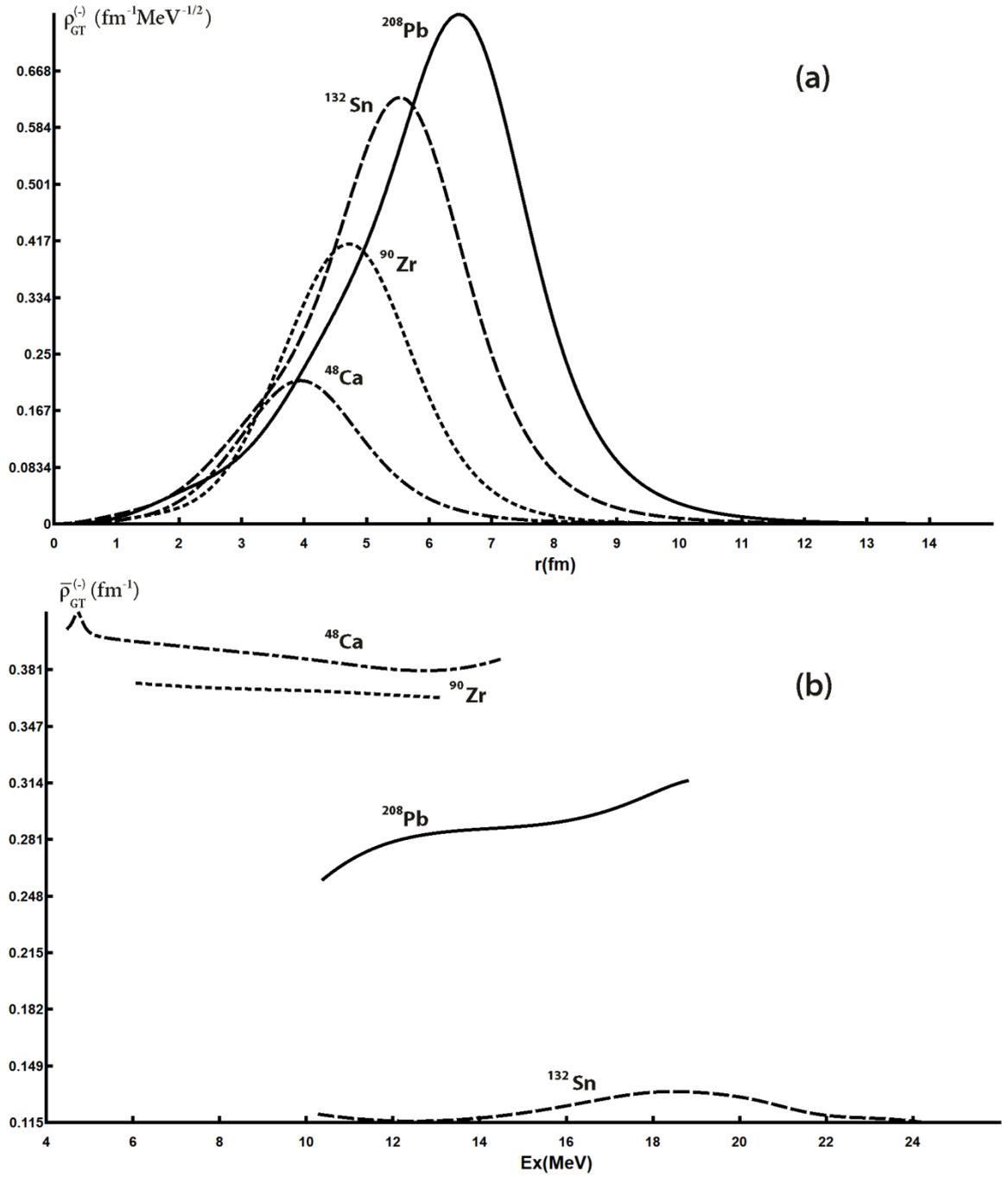

FIG. 4 The one-body (projected) transition density taken at the strength-function peak energy (a) and the reduced transition density $\bar{\rho}_V^{(-)}(r_{max}, E_x)$ taken at the maximum of $\rho_V^{(-)}(r, E_{x,peak})$ and shown in the excitation-energy interval $\delta_{max}^{(-)}$ (b). Both densities are calculated for GTR in parent nuclei under consideration.

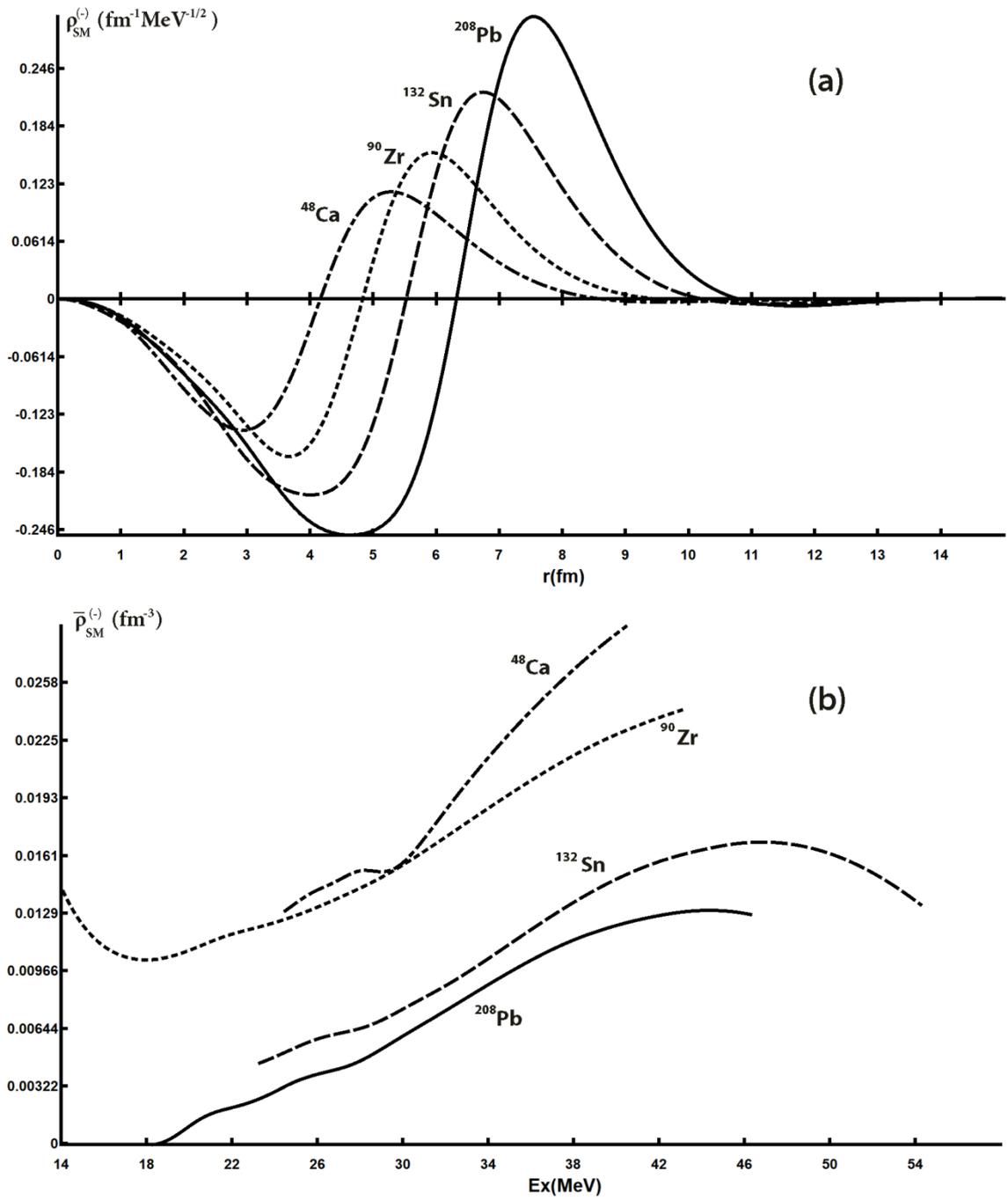

FIG. 5 The same as in Fig. 4, but for IVGSMR$^{(-)}$.

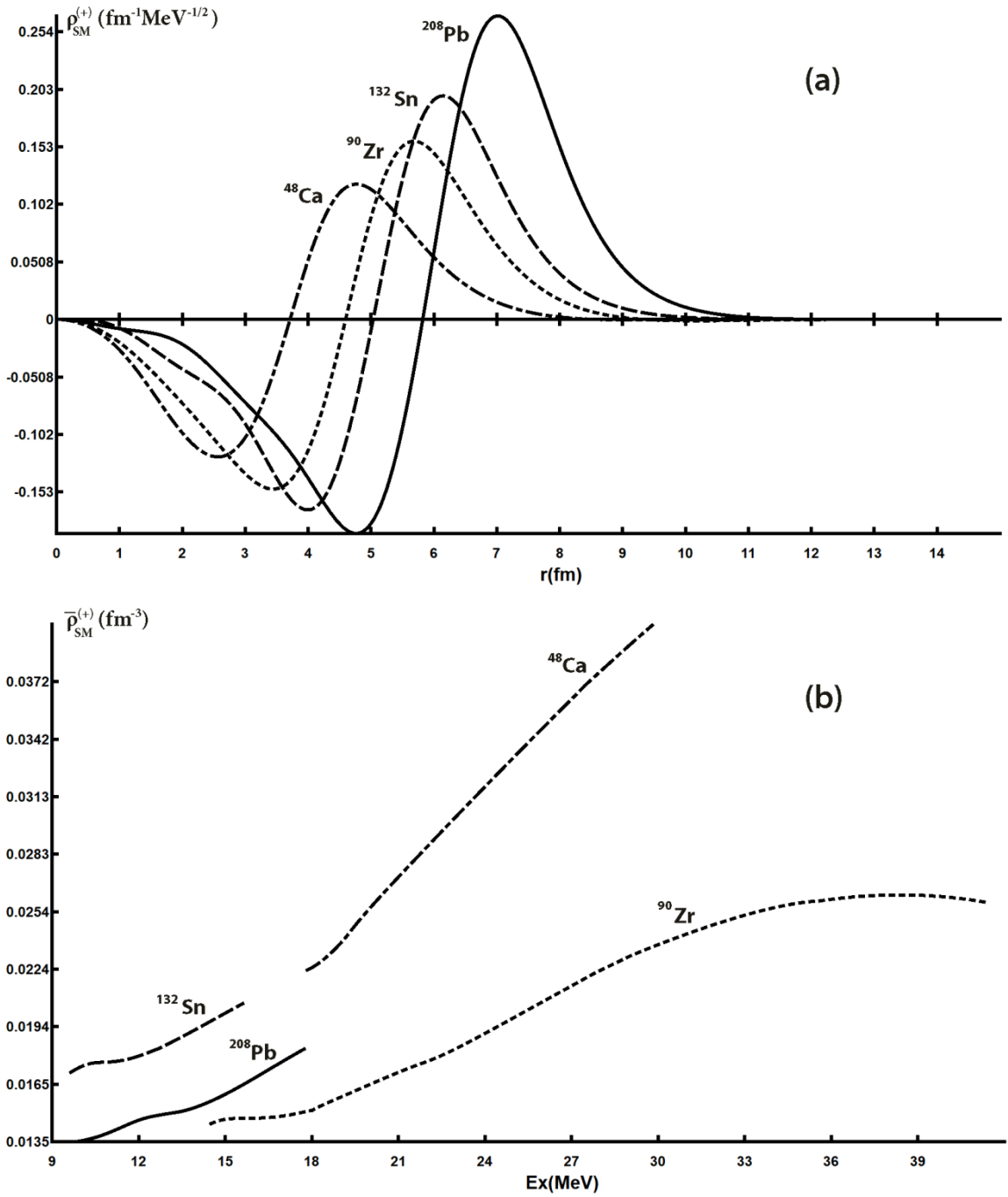

FIG. 6 The same as in Fig. 4, but for IVGSMR$^{(+)}$.

A possibility to evaluate the branching ratios for direct one-nucleon decay of GRs belongs to unique features of PHDOM. In Tables II – IV, we give the total branching ratios for direct one-nucleon decay of GTR, IVGSMR$^{(\mp)}$ in parent nuclei under consideration, calculated using Eqs. (6), (7), and (12). In the calculations, only decays from the excitation-energy interval $\delta_{max}^{(\mp)}$ (Tables II – IV) related to the GR main maximum are taken into account.

The calculated results given above demonstrate the ability of PHDOM in describing main characteristics and parameters of charge-exchange giant spin-flip monopole resonances in medium-heavy closed-shell parent nuclei. The Landau-Migdal parameter $g'$, the "spreading" parameters $\alpha$ and $B$ are adjusted to describe the observable peak energy and total width of GTR in each considered parent nucleus (Fig. 1, Tables I and II). Then, other characteristics and parameters of GTR (Fig. 4, Table II), the main properties of IVGSMR$^{(-)}$ and IVGSMR$^{(+)}$ (Figs. 2, 5, and 3, 6, Tables III, and IV) are described without using additional adjusted model parameters. A rather weak $A$-dependence of parameters $g'$ and $B$ should be noted. For nuclei from $^{90}$Zr to $^{208}$Pb, the values of Landau-Migdal parameter $g'$ used also in astrophysical applications are found to be close to the values deduced from an analysis of GTR excitation in direct charge-exchange reactions, $g' = 0.68 \pm 0.07$ [11]. All the strength-function calculations are verified by the use of the respective non-energy-weighted sum rules: the calculated fraction parameters x$^*$ are close to 100% (see Tables II – IV). As expected, the GTR$^{(+)}$ relative strength $x^{(+)*} = x^{(-)*} - x^*$ is very small for all nuclei under consideration (Table II). For this reason, the GTR total strength might be directly compared with the Ikeda sum rule. The relative strength of IVGSMR$^{(+)}$ is not-too-small for heavy nuclei and unexpectedly large for medium-mass nuclei (Table IV). The parameter $x_{max}^{(-)}$ calculated for GTR in $^{208}$Bi (Table II) is found in an acceptable agreement with the respective experimental value 60±15 [15]. It should be noted that this quantity, as well as the deduced $g'$ value, are slightly dependent on taking tensor correlations into account [9]. Concluding the description of GTR parameters, we note the reasonable description of the measured total branching ratio for direct one-proton decay of GTR in $^{208}$Bi, $b_{tot}^{(-),\uparrow} = 4.9 \pm 1.3$ % Ref. [15] (Table II). Experimental data concerned with the IVGSMR$^{(-)}$ parameters are available only for $^{208}$Bi: $E_{x,peak}^{exp} = 37 \pm 1$ MeV, $\Gamma = 14 \pm 3$ MeV and $b_{tot}^{(-),\uparrow} = 52 \pm 12\%$ Ref. [19]. The calculated peak energy is somewhat underestimated, while the total width and total one-proton direct-decay branching ratio are in an agreement with the experimental data (Table III). Reasons for theoretical underestimation of the peak energy and the missing of some main channels of direct one-proton decay of IVGSMR$^{(-)}$ in $^{208}$Bi in Ref. [19] are unclear at present.

Coming to the projected transition density taken at the peak energy of considered GRs, we note the expected node less transition-density radial dependence for GTR (main-tone GR, Fig.4(a)) and the one-node radial dependence for IVGSMR$^{(\mp)}$ (overtone GRs, Figs. 5(a), 6(a)). As follows from Figs 4(b) – 6(b), the use of the reduced projected transition density in describing nuclear reaction of GR excitation allows one to get directly information on the respective strength function.

## V. Summary and conclusive remarks

In this work, we present a description of main properties of Gamow-Teller and charge-exchange giant spin-monopole resonances in medium-heavy closed-shell parent nuclei. The description is realized within the semi-microscopic particle-hole dispersive optical model, which is a direct extension of the continuum-random-phase-approximation to taking into account (phenomenologically and in average over the energy) the spreading effect. Being "economic" in the practical use, the model allows us to describe for the mentioned resonances the strength function and projected transition density, both related to the appropriated probing operator, and probabilities of direct one-nucleon decay. Methods of verification of the calculated characteristic mentioned above are used within the model. As a rule, a reasonable description of the experimental data concerned with giant-resonance parameters is obtained. Extension of the model to taking into account tensor correlations in formation of the considered giant resonances in medium-mass closed-shell parent nuclei is in order.

## Acknowledgments.

The authors are thankful to M.L. Gorelik for his kind help in improving calculation algorithms and S. Shlomo for viewing the manuscript and valuable remarks.

This work is partially supported by Russian Foundation for Basic Research under Grant no. 19-02-00660 (V.I.B, M.H.U) and Program "Priority – 2030" for National Research Nuclear University "MEPHI" (M.H.U).